# The Mobile Devices and its Mobile Learning Usage Analysis

S. M. Jacob and B. Issac

*Abstract* — The usage of mobile devices for mobile learning is becoming increasingly popular. There is a new brand of students in the universities now-a-days who are easily connected to technology and innovative mobile devices. We attempt to do an analysis on a survey done with university students on mobile device usage for mobile learning purposes. This is to find the learning trends within the student community so that some of these popular practices could be encouraged to enhance learning among the student community. Both the quantitative and qualitative approaches are adopted in the analysis. The results are discussed and conclusions drawn in the end.

*Index Terms*—mobile learning, mobile devices, survey analysis, mobile applications

## I. INTRODUCTION

"Mobile learning is defined as the intersection between mobile computing (the application of small, portable and wireless computing and communication devices) and e-learning (learning facilitated and supported through the use of information and communications technology)", defined Clark Quinn, professor, author and expert in computer based education [1]. Or, mobile learning is learning that takes place with the help of mobile devices [2], [3].Wherever one looks, the evidence of mobile penetration is irrefutable: cell phones, PDAs, MP3 players, portable game devices, handhelds, tablets and laptops abound, reported Wagner [4]. It is a glaring truth that people of all walks and ages are increasingly connected and communicating to each other in ways that would have been impossible a few years ago. Naturally mobile computing integrated into e-learning make courses in the universities more accessible and portable. E-book versions are becoming as common as their printed counterparts. As companies work on new usability standards to the e-books, along with the addition of audio, video and text-to-speech components for e-book software, it would mark the widespread adoption of e-books without any barriers in the coming few years.

The paper addresses the fact whether the students in the particular university are ready to embrace mobile learning. Portable computing/communication devices such as laptops, PDAs, smart phones connected to wireless networks enable mobility and facilitate mobile learning. If properly facilitated, mobile learning can be of great benefit to learners by providing instructional materials and interaction through their mobile devices wherever and whenever they are on the move. It can be beneficial to the instructors also since they can access services and interact with students while on the move.

Naismith [5] had predicted the impact of emerging trends in mobile technologies on learning Compared to desktop PCs, restricted screen size and reduced interactivity are seen as the drawbacks of the mobile devices. But experiments have shown that despite limitations mobile media players offer a potential to increase learning that deserves further investigation [6].

The paper is organized as follows. Section 2 describes the common mobile computing devices and its usage in the university campus, section 3 shows some details of pedagogical implications of mobile learning, section 4 details on the survey done with the university students to measure the readiness, expectation of students concerning mobile learning and the extent of usage of mobile devices in the campus in the modern era of mobile computing. Section 5 portrays future challenges in m-learning, section 6 is on future mobile learning architecture, and section 7 is the conclusion The paper was also an effort on the part of the authors in response to the call made by Sharma and Kitchens [7] that it is high time that the instructors learn about and adapt to the changing environments and effectively facilitate mobile learning.

## II. MOBILE DEVICES AND ITS USAGE IN CAMPUS

Some of the common mobile computing devices and its features are detailed below for a brief analysis, in terms of its relevance to mobile learning [8].

*A. iPod* – iPod is a portable media player that allows a user to download music, podcasts, audio books and other video. Students can thus download lecture materials, audio and video lectures. With bigger screen iPods the users can read even e-books. The students can also share information files, work together on a project, provide visual directions or can interface with the iPod through a microphone. Pros: It helps in teaching support, as the professors in a university could give the audio or video lecture to the registered students as a free download. Cons: The cost can be a factor where all the students cannot afford to use one. It also does not provide interactivity and the screen size is generally small to read large chunks of data.

*B. Personal Digital Assistant (PDA)* – PDAs form a good combination of digital storage along with computing power,

internet access, wireless network access through Wi-Fi or Bluetooth, and pen or stylus input interface, along with other word processing tools. It lets users access email and web content and can play audio and video files. It supports interactive and group learning. Pros: Since text and data entry is possible through the screen keyboard or stylus, PDA with its relatively large screen, stands as a favorite choice since it integrates communication tools in it. Cons: It may be slightly bulky for a normal sized pocket.

*C. Smart Phone* – Smart phone integrates telephone features, along with camera, PDA and MP3 player. It also supports access to Internet. Users can download audio or video lectures, flash movies, edit text documents, send IM and use the phone for storing data. It supports interactive learning as it enables global collaboration. Pros: It combines a host of options and features in one easily portable device. Cons: The issue of small screen makes reading the text and web browsing quite difficult. The cost of some advanced smart phones is quite high.

*D. Laptop or Tablet PC* – Laptop or tablet PC is the most functional of all the mobile devices and it has all the features of a workstation PC. It comes with the network support for Bluetooth, WI-Fi and Ethernet. Tablet PCs also integrate handwriting recognition, voice to text conversion etc for input. These computing nodes could support email, web surfing, word processing, Instant Messaging, VoIP connections and many other application programs. Lot of interactivity and collaboration in research can be thus supported. Pros: The laptop provides the most powerful computing environment with mobile devices. Cons: The relatively large size and lack of mobility-on-the-run limits its network usage, where mobile network services are available.

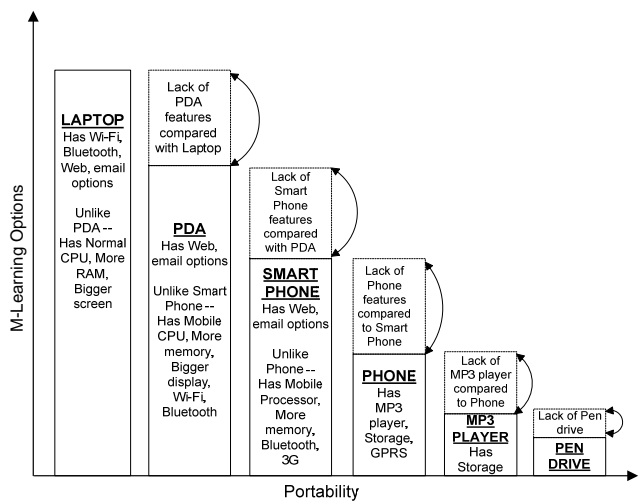

Figure 1. The Portability versus M-Learning options of various mobile devices, as a graph based block diagram.

*E. Others: MP3 Player, Pen Drive (USB Drive), Hand phone* – MP3 player is a digital audio player which plays music and audio files, but no interactivity is offered. This could be used by students to listen to podcasts and audio lectures. Pros: Being light and compact, they have good audio quality output and lasting battery life. Cons: No interactivity options are available and it can be replaced by other hearing gadgets or options. The Pen Drive is a mass storage device which attaches easily to many computers and other devices. The portable drive finds great use by students to transport files easily between university and home. Pros: Being portable and handy, it proves to be a good medium to carry files while on the move. Cons: Other devices with integrated storage features can replace pen drives for mass storage. A regular hand phone has telephone and other added facilities, except for the full PDA features in comparison to smart phones. Pros: It is more affordable than a PDA and Smart phone and has many integrated features for audio and storage. Cons: It lacks the processing ability in comparison to PDA that aids mobile learning.

The comparison of portability versus m-learning options with these various devices can be shown as in fig. 1. Laptop and PDA tops in features that support mobile learning. Here each device is compared with the device listed or shown as a bar just before it on the left side, showing how much it lacks in its features in comparison.

III. PEDAGOGICAL IMPLICATIONS OF MOBILE LEARNING

In order to understand how people learn, the progression of behaviorism, cognitive science and constructivism serves as the foundation in guiding the migration of implementing technologies in learning and, hence mobile learning too. Mobile learning could be adapted to the following types of learning by the instructors.

*A. Learning Types*

*Behaviorism* – This propounds feedback and reinforcement. Mobile devices can facilitate these when faculty and students are using the devices in tandem.

*Constructivism* – This demands rich media, simulations and immersive environments. Simulations, visualization and gaming environments could be provided through mobile devices at the convenience of students.

*Informal or situated learning* – This talks of using education in "context aware" environments relevant to the field of study. Mobile devices allow content portability into "context aware" environments.

*Collaborative learning* – This propagates recording and sharing instantly. With the handy and portable mobile devices, there are much possibilities of creating and sharing student and teacher authored resources.

It is true that there is a developing trend in information technologies that provide interactive mechanism among the learners, instructors and the learning material. But effective learning could happen only when the learner decides to engage himself actively and cognitively in the learning activities.

*B. Attractive Learning Environments*

The flexibility, instant connectivity, mobility associated with mobile learning has given rise to new delivery platforms for teaching and learning. Below are given some of the improvements in the learning environment:

(1) Both faculty and staff have access to ubiquitous learning environments.

(2) Students have access to the same hardware and software as faculty.

(3) Cost savings result in replacing desktop computers with laptops and hard wired networks with wireless ones.

(4) Standardization of platforms maximizes access and minimizes need for technical support [9].

IV. ANALYSIS OF STUDENT SURVEY

*A. Survey Design*

To determine the extent of usage of mobile devices in the campus, an initial survey similar to the one in [8] was done among a random population of 151 undergraduate students in a Malaysian university. The response rate was 33%. The objectives of the survey were to check on: (1) students' appreciation towards mobile learning and the popularity of the laptop; (2) the extent of usage of the different mobile devices and the kind of mobile activities engaged in; (3) and the possibility of bringing in smaller mobile devices like smart phones or PDAs in mobile learning; (4) their expectation regarding when true mobile learning would happen; (5) the reasons for supporting mobile learning.

The questionnaire was intended to collect quantitative and qualitative data. In view of the first objective, three statements were given to measure student responses on a 5-point Likert scale from 1-5, where 1 represents "Strongly Agree" (SA), 2 represents "Agree" (A), 3 represents "Unsure" (US), 4 represents "Disagree"(D),5 represents "Strongly Disagree"(SDA). They were: (1) Mobile learning should be supported in this mobile, digital era. (2) Mobile learning is an innovation in education. (3) Among the listed mobile devices - iPod, MP3 player, PDA, Pen Drive, Cell phone, Smart phone and laptop, I rank laptop to be the best/efficient form of mobile device for mobile learning, available now.

In alignment with the second and third objectives, there were given three questions for which multiple answers could be chosen. (1) "Which of the following mobile devices do you own?" The choices were given as – iPod, MP3 player, PDA, Pen Drive, Cell phone, Smart phone, Laptop. (2) "Devices like mobile/smart phones or PDAs can be very popular learning tools if the following technology issues can be addressed well. Choose the relevant options that you feel are a must." The choices given were: Larger displays/screens, Lower network traffic, Larger memory capacities, Faster data transmission, Technology unification (integration of wireless technologies like 3G, Wi-Fi, WiMAX, bluetooth etc), and Better proliferation/deployment/implementation. (3) "What mobile activities do you do?" The choices given were: Download and read documents, Send and receive emails, Send and receive instant messages, Send and receive short text messages, Send and receive Multimedia messages, Transfer files from one place to another through pen drive, Play interactive games through internet or through hand held game devices, Transfer photo/audio or other data through hand phones or smart phones.

The fourth objective was based on a hypothesis proposed by the authors that the true mobile learning would happen in less than the next five years. A question was asked: When will true mobile learning happen? The choices given were: in the next one year, in the next three years, in the next five years, in the next ten years, it will never happen; it's only a dream.

Lastly, in line with the fifth objective, open comments were invited with at least two supportive reasons to support or not support the fact that mobile learning is an alternate way of learning. This represented the qualitative part of the data collection.

*B. Survey Results — Quantitative Analysis*

Table 1 show the responses to the three questions whose responses were measured using the Likert scale. 66% of the students were agreeing to the fact that mobile learning should be supported in this mobile era. The mean response was 2.2 with a standard deviation (SD) of 0.9. Both values point to a consistency in agreement to the statement. 84% were of the opinion that mobile learning is an innovation in education. The respective mean response value of 2.1 and standard deviation (SD) of 0.7 confirms the majority of students are in agreement with the statement. Nearly 90% agreed that laptop is the best/efficient form of mobile device available now. This confirms the popularity of laptops in mobile learning [10]. The mean response here is 1.7 with a standard deviation (SD) of 0.8, which point to a strong agreement in favour of the statement.

TABLE I  RESPONSES OF UNIVERSITY STUDENTS

| Items | Responses | | | | | | |
|---|---|---|---|---|---|---|---|
| | SA | A | US | DA | SDA | Mean | SD |
| Mobile learning should be supported in this mobile, digital era | 10 | 23 | 14 | 2 | 1 | 2.2 | 0.9 |
| Mobile learning is an innovation in education | 7 | 35 | 6 | 1 | 1 | 2.1 | 0.7 |
| Laptop is the best/efficient form of mobile device available now | 24 | 20 | 2 | 3 | 0 | 1.7 | 0.8 |

Fig. 2 reveals that the most popular devices among students are pen drives, cell phones and laptops. The users of MP3 players are the next highest category followed by iPods and smart phones. PDAs seem to be the least used mobile device in the list. The findings are in line with the results in [8].

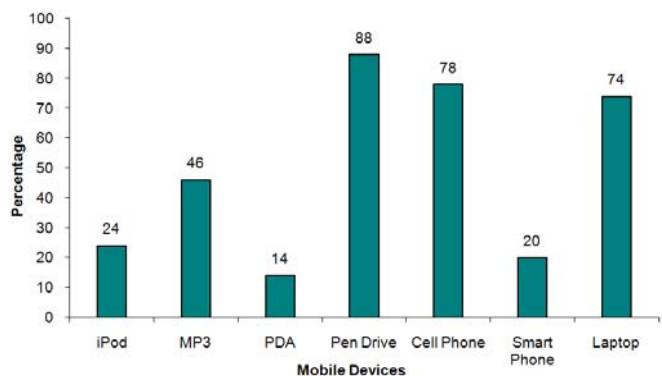

Figure 2. The usage of mobile devices among students.

Table 2 shows the interest of the students in the technology issues to be addressed. More than 50% responses were evident in relation to larger displays, larger memory capacities, faster data transmission and technology unification. Faster data transmission and technology unification seem to have bothered the students more than the others.

*Hypothesis: True mobile learning will happen in the next 3-5 years.* Fig. 3 reveals the hypothesis is true from the point of view of the students too. A majority of about 75% have predicted that this would take place within the next five years or less, with the highest percentage of about 40% supporting the technology would be fully active in the next three years. The responses were indicative of the readiness for mobile learning in the campus.

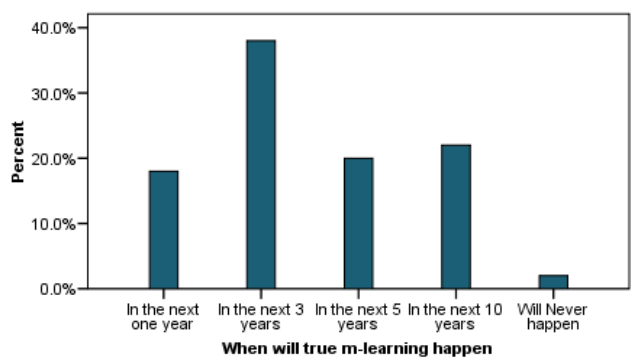

Figure 3. The expectation of true mobile learning to happen.

TABLE II   TECHNOLOGY ISSUES TO BE ADDRESSED IN ORDER TO MAKE SMALLER MOBILE DEVICES POPULAR

| Technology issues to be addressed | Percentage |
| --- | --- |
| Mobile devices must have larger displays/screens | 54 |
| Mobile devices must have lower network traffic. | 30 |
| Mobile devices must have larger memory capacities. | 52 |
| Mobile devices must have faster data transmission. | 64 |
| Mobile devices must have technology unification. | 64 |
| Mobile devices must have better proliferation/deployment/implementation | 38 |

Among the kind of mobile activities engaged in by the students, Table 3 shows that sending short text messages, transferring files using the pen drive and email usage top the list. This undoubtedly confirms the modern day trend of high usage of mobile phones, pen drives and emailing via laptops. Still, 50% or more do engage in the other activities listed in the table which shows the popularity of the usage of mobile devices. The findings of table 3 tallies with the results of fig. 2 which also shows the popularity of mobile phones, pen drives and laptops among the different mobile devices owned by the students. It looked as if many of these students were living in a digital world.

TABLE III   TYPE OF MOBILE ACTIVITIES THAT STUDENTS ENGAGE THEMSELVES IN

| Type of Mobile activities engaged in | Percentage |
| --- | --- |
| Send and receive emails | 82 |
| Send and receive instant messages | 68 |
| Send and receive short text messages | 88 |
| Send and receive Multimedia messages | 50 |
| Transfer files from one place to another through pen drive. | 84 |
| Play interactive games through internet or through hand held game devices | 58 |
| Transfer photo/audio or other data through hand phones or smart phones | 60 |

*C.  Survey Results — Qualitative Analysis*

The qualitative comments given by students were in response to the statement: "Mobile learning is an alternate way of learning – Give two reasons to support or not support this." Most of the comments were supportive and we are giving a sample of both type of comments.

- *Supportive Comments:*

1. I support mobile learning as an alternate way of learning. This is because the contents can be retrieved more easily. It also promotes learning without geographical limitations.

2. It is widely used and it has become a need nowadays (although it is still a luxury). We have now entered into an era of technology. We will drop out if we are not updated or not learned.

3. Because if some students don't attend the class, they still can download the audio file that have been provided and will not miss the topic taught in class. It also can be used by those students who miss some parts in the lecture class.

4. With mobile learning, I think I could learn anytime and anywhere without bringing bulk of books or materials around.

5. If all these devices such as hand phone, PDAs and laptop integrated together....mobile learning is fun..

6. I support mobile learning indeed. With mobile learning, students can just repeat the lecture again and again using their iPod or they can save the lecture slides to their PDA so that they can refer back anytime.

7. Mobile learning should be encouraged in the future years, mainly as a way to support our legacy lecture system. I don't mean that attending lecture class should be diminished. Mobile learning should be considered as a way to support teaching in the class, e.g. as a learning supplement.

8. This will definitely make the subject much more interesting to learn. It's a new way to learn, so this might help us to expand our knowledge and live towards the new technology. Hence I support this.

- *Non-Supportive Comments:*

1. I don't support, simply because not every student owns sophisticated electronic gadgets since they are expensive and buying those things doesn't really support studies. In contrary, those electronic gadgets cause students to fool around and entertain themselves.

2. I don't support especially because for teenagers, having classes virtually (e-learning or virtual world classes) could be a problem. Not only that, students tend to get easily distracted, it would be even worse when they are somewhere doing something they prefer than to "attend class".

3. Given the slow development rate of our state, its very unlikely that this feature will be accomplished any time soon, could take ages. Paper documents/ printed out docs are the most efficient for me. They help me to concentrate better when looking at the paper than at a screen.

4. I personally think online media like Blackboard is sufficient enough for learning (no need for IPod, PDA etc).Otherwise learning from textbook is somehow more efficient, because by reading we will get detailed explanation rather than searching anywhere else on the Web because what is available on the net is too brief and not too trustable since everyone can manipulate the information on the Web.

5. It can only act as an added support in learning. It can never be a substitute for classroom learning. The interaction between students-students and teacher-students goes a long way in learning anything.

*D. Summary of Survey Results*

1. The readiness of the students for full fledged mobile learning was very clear.

2. A majority expected true mobile learning to be in vogue within the next 3-5 years.

3. The popular mobile devices used among the students were pen drives, cell phones and laptops. The most popular mobile activities correspond to these devices as well – send and receive short text messages, send and receive emails, transfer files through pen drive.

4. The students raised technology issues like faster data transmission, technology unification of mobile devices, larger display screens and larger memory capacities to be addressed in order to make mobile devices popular.

5. The supportive comments show the students' enthusiasm for podcasting, technology unification of mobile devices. They see m-learning as a means to make the subject interesting and an effective learning supplement.

6. The non supportive comments convey the fear that technology would fully replace the direct teacher-student interaction; non affordability of mobile devices to some students; technology being a distracter to serious study time.

Given the overall positive feedbacks and balanced negative feedbacks, the authors conclude there is a great potential of introducing mobile technologies into the teaching and learning process of higher education. These feedbacks are in line with the study done in [11]. The study provides the foundation for the further development and expansion of mobile devices integration in higher education.

V. FUTURE CHALLENGES OF M-LEARNING

There are quite a number of challenges with m-learning when we consider all the available mobile devices and their contribution toward m-learning. The major ones are as follows:

*Adaptive Learning* – This demands that the instructional strategies and learning content should be designed to adapt to the learner's profile and personal needs. Hence to effect adaptive learning, the learners' location should be taken into consideration.

*Limited Text Display* – There needs to be explored how mobile devices could support in providing continuous learning activity during the learning courses, or a stand alone learning module.

*Instant Communication* – Location and response time are crucial factors in supporting the success of good academic interaction and learner satisfaction. Prompt interaction among learning peers could be built in by the mobile communication network by utilizing the prompt notifications of message reception [12].

VI. FUTURE M-LEARNING ARCHITECTURE

The authors foresee a future convergence network that could possibly bring about learning on the move without restrictions. For a university or school, it may be good to support the existing teaching methods with an m-learning infrastructure that aids learning anytime and anywhere. In such a scenario, many wireless technologies like Wi-Fi, 3G, Wi-Max, Bluetooth etc. have to integrate into one convergence network, where the technology runs in the background and the user is not aware of the changes in technology or transition from one network to another as in fig. 4. As of now, this 'dream' convergence of different wireless networks are still in research with the possible emergence of 4G networks in future [13]. The survey from table 2 is a clear indication that students look forward to technology unification, in order to make smaller mobile devices as popular learning tools.

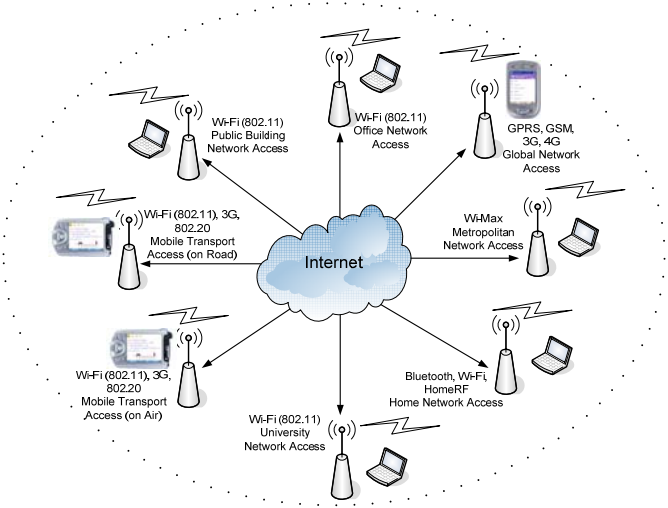

Figure 4. The future convergence network for mobile learning [13].

VII. CONCLUSION

Mobile devices and its associated technologies are necessary components for the sustenance and growth of mobile learning. As the learner population gets access to these technologies, the practitioners of mobile learning also need to get themselves acquainted well with the same. The efforts of the authors in looking into the options of mobile device technologies available for mobile learning were fruitful. The quantitative and qualitative analyses of the student survey revealed the mobile device usage among university students. The authors are resounding the fact that the goal of mobile learning is to facilitate learning and, in future, wireless convergence networks could offer ubiquitous anytime and anywhere learning a ground reality.